# Matter: IoT Interoperability for Smart Homes

Saeid Madadi-Barough, Pau Ruiz-Blanco, Jiadeng Lin, Rafael Vidal, Carles Gomez
Universitat Politècnica de Catalunya

*Abstract*—**The smart home is a major Internet of Things (IoT) application domain with tremendous market expectations. However, communication solutions for smart home devices have exhibited a lack of interoperability, challenging the success of the smart home concept. Aiming to overcome this problem, crucial industry organizations have collaborated to produce Matter, a connectivity solution intended as a universal smart home standard. This paper overviews, evaluates and discusses Matter, focusing on its design, features, performance, and future directions.**

## I. INTRODUCTION

The smart home is a major application domain of the Internet of Things (IoT). In the smart home vision, home objects are equipped with connected, inexpensive machines that often include sensors and actuators. In this paradigm, information collected from the home is analyzed, leading to actions that allow efficient resource management and enhanced user comfort. The smart home market is expected to steadily increase, reaching USD 207 billion worldwide in 2026 [1].

The vast potential of the smart home concept has attracted the interest of the industry, academia and standards development organizations for decades [2]. A wide diversity of communication solutions have been used in the smart home. However, the heterogeneity of such solutions challenges device interoperability, complicates product development, and threatens user adoption of smart home technology. For example, newly acquired devices will not integrate into a user's pre-existing home network if they cannot interoperate with the latter. Market fragmentation limits the range of products the user can consider, and increases their cost.

Many smart home protocol stacks share no common component. Interoperability between devices implementing such different protocol stacks is only possible via a protocol translation element, often called a *hub*. On the other hand, two decades of IETF standardization efforts have produced IPv6 support over several IoT technologies [3]. Therefore, IPv6-based devices using different IoT technologies can at least interoperate at the network layer via an IPv6 router, and can also support end-to-end (E2E) IP-based Internet connectivity. However, common functionality is still needed at the higher layers of the protocol stack: several IoT application-layer protocols, along with different data models, message formats, and interaction methods, exist in the market. Consequently, a common standard for smart home devices is needed.

Aiming to produce a new connectivity standard to achieve smart home interoperability, a working group called Project Connected Home over IP (PCHIP) was formed in 2019. PCHIP leaders included Google, Amazon, Apple, and the ZigBee Alliance. The latter recently rebranded as Connectivity Standards Alliance (CSA), now including the former PCHIP members. In late 2022, CSA released the new smart home connectivity standard, called Matter. The standard's name alludes to its universal purpose, since *matter* is common to all physical objects. The current Matter version is 1.3 [4].

In contrast with other smart home solutions (e.g., ZigBee, Z-Wave or Bluetooth Mesh), Matter is an open, IPv6-based protocol stack that introduces new eponymous application-layer functionality over several underlying technologies, such as Wi-Fi, Thread or Ethernet. Matter is expected to significantly impact the smart home as a unifying standard for several billion devices.

While the currently scarce literature on Matter concerns rather its market adoption [5], in this paper we focus on its design, features, and performance as a networking solution.

The remainder of this paper offers the following contributions: i) a holistic, tutorial-style Matter overview (Sections II-III), ii) the first ever Matter performance evaluation (Section IV), iii) a comparison of Matter with other prominent smart home solutions (Section V), iv) an interoperability cost analysis (Section VI), and v) a discussion of future directions (Section VII). Section VIII concludes the paper.

## II. MATTER FUNDAMENTAL CONCEPTS

This section introduces Matter fundamental concepts, including the term called fabric, the supported network topologies, and the related interconnection devices.

A device in a Matter network belongs to one or more fabrics. A *fabric* is a security domain that comprises a set of nodes that share a common root of trust. Data transmission from one node to one or more destination nodes occurs within a fabric. A node is added to a fabric and configured by means of a process named commissioning.

During commissioning, Matter employs Public Key Infrastructure (PKI) for device attestation, which allows a commissioner to assess whether the joining node is authentic. To this end, a Matter device is shipped with a unique Device Attestation Certificate (DAC), which is signed by an entity appointed by CSA to act as Certificate Authority (CA) or by an authorized intermediate. After successful DAC verification, PKI is also used to trustworthily supply the node with credentials. These include a 64-bit identifier called node ID, and a Node Operational Certificate (NOC). The latter binds a unique node key pair to the node ID. The NOC is signed by the commissioner, acting as CA and root of trust for the fabric.

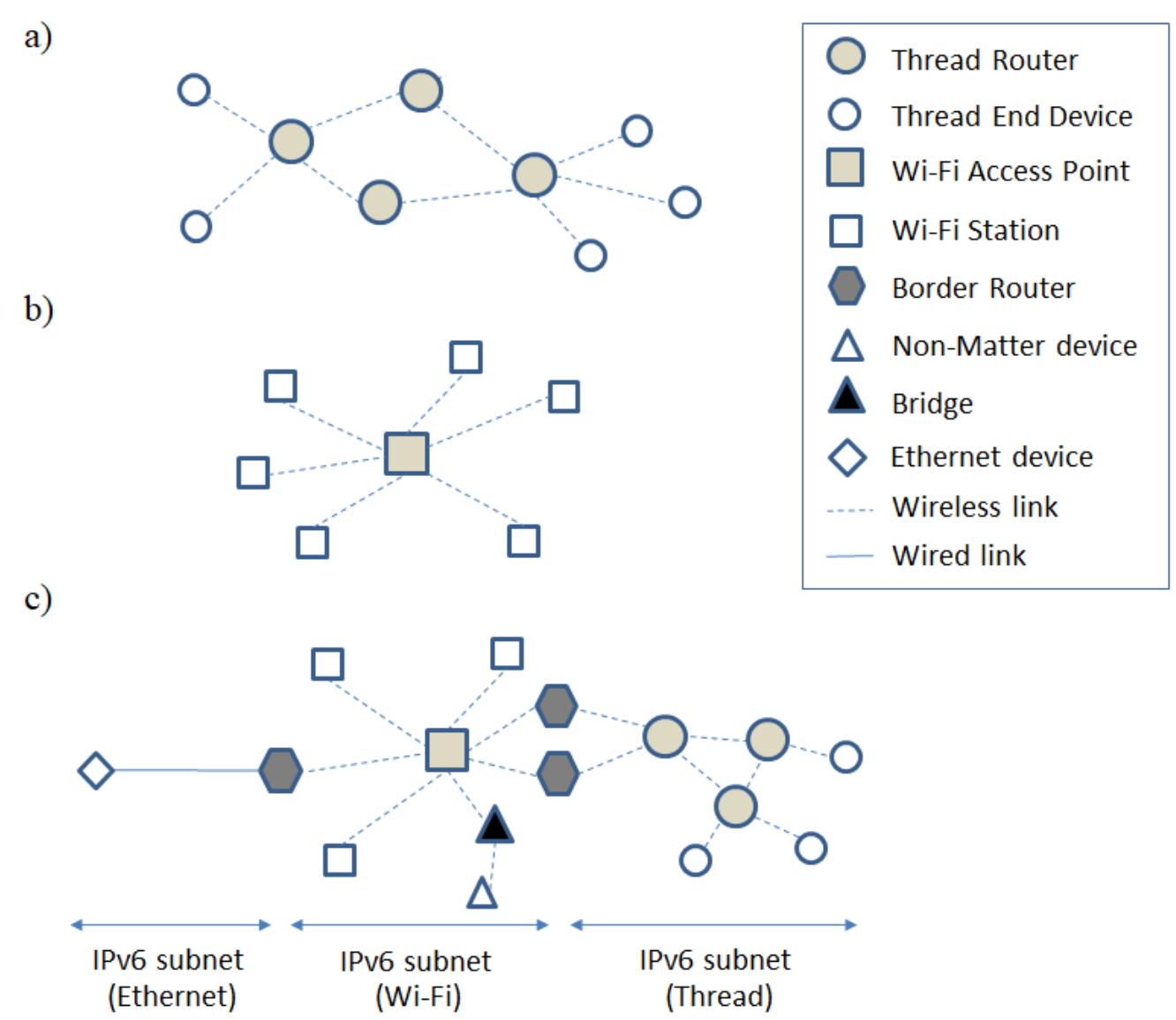

Fig. 1. Matter network topology examples: a) Thread single network topology, b) Wi-Fi single network topology, c) star network topology, with a central Wi-Fi network, and two peripheral networks (Ethernet and Thread).

A fabric may be supported by two types of network topologies: the single network topology, and the star network topology (Fig. 1). The former comprises a collection of nodes using the same technology, whereas in the latter there is one central hub network (usually, the main home network, based on Wi-Fi or Ethernet) connected to a number of peripheral networks of any supported kind (e.g., Thread, Wi-Fi or Ethernet). A peripheral network is connected to the central network via one or more Border Routers. Each central or peripheral network is an IPv6 subnet. Non-Matter devices can also participate in a fabric via a Bridge. A Matter network can either be connected to the Internet or isolated.

## III. Matter Protocol Stack

The Matter protocol stack has been conceived to enable interoperability for smart home devices that use different underlying technologies. Many such devices are characterized by significant constraints in processing and memory resources. Accordingly, Matter has been designed to be suitable for devices with 128 kB of RAM and 1 MB of Flash memory. Furthermore, some smart home devices rely on a limited energy source, such as a simple battery. Accordingly, Matter supports low energy operation.

Matter comprises four main layers (Fig. 2): the application layer, the transport layer, the network layer and the technology layer. The application layer represents the main novelty introduced by Matter, whereas underlying layer protocols have been selected to provide suitable E2E functionality, network-layer interoperability, and support for main smart home technologies. Note that Bluetooth Low Energy (BLE) is only used for commissioning. The next subsections present the funcionality of each layer, by following a top-down approach.

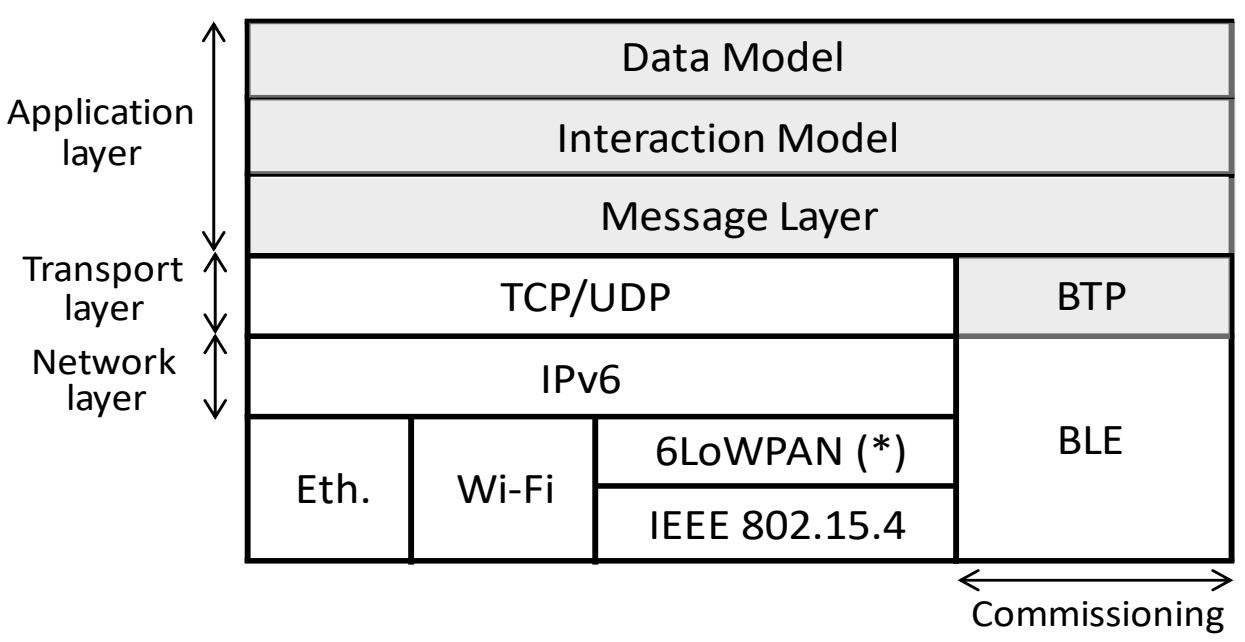

Fig. 2. Matter protocol stack. Shaded components are defined by the Matter specification. Thread is an IPv6-based solution that exploits 6LoWPAN to operate over IEEE 802.15.4. (*) Includes Thread routing.

### *A. Application layer*

Matter defines a new application-layer protocol that comprises three functionality sublayers, namely: the Data Model, the Interaction Model, and the Message Layer (Fig. 2). These sublayers are overviewed next.

#### *1) Data Model*

Applications and network operation itself require handling various data types. To ensure interoperability and to ease implementation, such data types, and their properties, need to be standardized and efficiently organized. The Data Model defines the data types, formats, qualities, and structures used in Matter, based on data constructs called elements. An *element* is characterized by an identifier, a name, access features (e.g., read, write, invoke, and related privileges), and qualities such as data type, associated responses or direction. Example elements include fabric, node, endpoint, cluster, command, event, or attribute.

A *node* is a physical or virtual object with a set of capabilities (e.g., a thermostat). A node comprises at least one entity called endpoint which provides a service (e.g., a thermostat's temperature sensor).

A *cluster* defines a client and a server that communicate to achieve a purpose via interactions. The cluster server manages elements (e.g., attributes, events, and commands), whereas the cluster client initiates interactions to manipulate server elements.

An *attribute* is a data structure with standardized metadata that define qualities of such data and associated behaviors. An *event* corresponds to registered data on something that has occurred. A *command* is a data unit aimed at producing a behavior on its receiver.

Matter clusters have been defined in the application areas of measurement and sensing, lighting, HVAC, closures, media, robots, and home appliances. There are also mandatory clusters that support network management, such as Over-the-Air (OTA) software updates. The latter may be needed for security or functionality reasons. In Matter, an OTA Requestor obtains software updates from an OTA Provider. The former discovers OTA Providers by using records set during commissioning or via OTA Provider announcements. The OTA Requestor selects an OTA Provider. An updated software image is transferred to the OTA Requestor by using Matter's Bulk Data Exchange protocol. Thereafter, the OTA Requestor polls the OTA Provider periodically for new software updates.

The Data Model design is based on the ZigBee Cluster Library (ZCL), which provides similar functionality in the ZigBee protocol stack. ZCL messages were devised so that most of them fit a single 127-byte IEEE 802.15.4 frame (i.e., the usual maximum-sized link-layer frame in ZigBee). Data Model messages inherit such size features, and are suitable over Thread (which is also based on IEEE 802.15.4, see subsection III-D.2).

#### 2) *Interaction Model*

The Interaction Model specifies how two or more endpoints may communicate, by using predefined sequences of message exchanges organized hierarchically into so-called interactions, transactions, and actions. An *interaction* comprises one or more transactions, whereas a *transaction* comprises one or more actions. An *action* is an elementary communication that requires the transmission of one or more messages from one origin endpoint to one or several destination endpoints.

There are four types of interactions, namely: Read, Write, Invoke, and Subscribe. The first three are composed of a single transaction each (called Read, Write, and Invoke, respectively), whereas the latter comprises two transactions (called Subscribe and Report). Read allows to obtain attribute or event data, Write aims to modify attribute data, and Invoke solicits commands on target nodes. Subscribe allows a node to receive attribute or event data reports from a *publisher*. In the Subscribe transaction, a *subscriber* first sends a subscribe request to a publisher. The latter replies with a data report, which contains the current value of the data of interest to the subscriber. Subsequently, the publisher starts the Report transaction, where it reports data to the subscriber whenever the target data value changes or after a maximum interval when such change does not occur.

There exist three types of actions: request, response and report. Each interaction involves at least a pair of consecutive actions which are semantically related (e.g., a request and its response).

The Interaction Model also defines the concept called path, which identifies a cluster attribute, event or command on a given target endpoint. Paths are included in the messages that convey actions on corresponding elements.

Each action is usually encoded into one message, although an action handling long payloads such as lists can be carried by a set of chunked messages. Message format and encoding are a responsibility of the underlying Message Layer.

#### 3) *Message Layer*

The Message Layer, which is located atop the transport layer, performs secure E2E transmission of actions. To this aim, it defines data units called messages, which may carry application data (i.e., actions) or control data. An action is encoded by means of message header fields that identify the action, the transaction it belongs to, and an optional payload.

All application data exchanged is secured by means of an underlying session that provides message encryption, authentication, and integrity. The session is created by using the Certificate Authenticated Session Establishment (CASE), prior to application data transmission. CASE allows the secure exchange of NOCs and other materials to establish a shared symmetric encryption key between the involved endpoints, and for their mutual authentication.

A session may multiplex several concurrent transactions. Each message is prepended with header fields that identify the corresponding session and details on the security services used. A 16-byte Message Integrity Check (MIC) field is appended to each message. The Advanced Encryption Standard (AES) block cipher in Counter with Cipher block chaining Message authentication code (CCM) Mode (AES-CCM), with a 16-byte key, is used to generate the MIC and to encrypt it along with the message payload. The 16-byte MIC length is more aligned with the one generally used on the Internet (e.g., with TLS or DTLS) than for constrained devices, where an 8-byte MIC is currently state-of-the-art [6]. This choice provides exponentially stronger message integrity protection, at the expense of increasing transmission and processing overhead, producing non-negligible performance impact for devices with constrained CPU, memory, transmission rate, and energy.

The Message Layer also offers the Message Reliability Protocol (MRP), which provides a simple, optional, and per-message E2E reliability mechanism that may be used atop UDP (see the next subsection). In such case, when a sender needs to transmit a message reliably, it activates a flag of the message header and triggers an Automatic Repeat reQuest (ARQ) mechanism with positive Acknowldgments (ACKs) and exponential backoff. Duplicate messages (produced by either MRP or lower layer retries) are ignored by a receiver.

There are two types of MRP ACKs: standalone ACKs and piggybacked ACKs. In the latter, a data message also carries the ACK. To exploit opportunities for piggybacking ACKs, a receiver defers ACK transmission.

### B. *Transport layer*

As part of Matter's overarching goal of interoperability, the classic transport-layer protocols of the TCP/IP stack, TCP and UDP, can carry Matter messages. Matter also offers the Bluetooth Transport Protocol (BTP), a reliable transport-layer protocol solely used for commissioning over BLE (Fig. 2).

TCP has often been deemed inadequate for IoT scenarios, after claims regarding its potential complexity, protocol overhead, and underperformance in wireless environments. Accordingly, early IP-based protocol stacks for IoT resorted to using UDP, along with optional and simple (e.g., stop-and-wait) E2E reliability at the application layer. One such example is the original design of the Constrained Application Protocol (CoAP) [7]. Matter supports a similar approach, offering MRP over UDP.

However, TCP is needed in some IoT environments, such as networks including UDP-unfriendly Network Address Translation (NAT) middleboxes. Furthermore, it has been shown that most claims against TCP in an IoT context are not valid or fair [8]. Matter supports both TCP and UDP, offering adaptability to each particular scenario.

### C. Network layer

At the network layer, and as its cornerstone for Internet connectivity and device interoperability, Matter uses the Internet Protocol version 6 (IPv6). In fact, due to the Internet's success, available IPv4 addresses are currently almost exhausted. In contrast, IPv6 offers a virtually unlimited address space, which suits the need to deploy billions of IoT devices, and it also provides tools for unattended network operation.

### D. Technology layer

To ensure proper provisioning functionality and manageable certification workload, the initial Matter version limits its supported lower-layer technologies to Ethernet, Wi-Fi and Thread. In this subsection, we focus on the last two, since they are relevant technologies from an IoT perspective, whereas Ethernet is rather intended for network infrastructure.

#### 1) Wi-Fi

Wi-Fi is the brand name for certified devices that implement the IEEE 802.11 standard for Wireless Local Area Networks (WLAN). This standard defines physical- and link-layer functionality, using Carrier Sense Multiple Access/Collision Avoidance (CSMA/CA) for medium access.

Since Wi-Fi was not originally designed for IoT, it is generally not optimized for energy performance. However, many Wi-Fi smart home devices are mains-powered, therefore they have a virtually unlimited energy supply.

Wi-Fi is typically deployed as a star network topology, where devices are connected to a Wi-Fi router. Since coverage may be insufficient for a whole home space, range extenders or mesh solutions are gaining market presence. In contrast, smart home IoT technologies natively support the mesh topology [2].

Wi-Fi naturally supports the IPv6 Maximum Transmission Unit (MTU) requirement, whereby the layer below IPv6 must be able to handle packets of at least 1280 bytes, since its basic maximum frame payload size is 2304 bytes.

#### 2) Thread

Thread is an IPv6-based solution designed for smart homes that operates over an IEEE 802.15.4 mesh network [9].

IEEE 802.15.4 was the first open physical- and link-layer radio standard designed to enable applications for simple devices with significant energy constraints and relaxed bandwidth requirements (e.g., sensors). IEEE 802.15.4 has been developed as a Low-Rate Wireless Personal Area Network (LR-WPAN) technology. IEEE 802.15.4 supports the star and the mesh network topologies. It has become an IoT cornerstone, providing the lower layers for IoT protocol stacks such as ZigBee, ISA 100.11a, 6TiSCH, Wi-SUN, and Thread.

Thread uses the IEEE 802.15.4 2.4 GHz band, which offers a bitrate of 250 kbit/s. For medium access, IEEE 802.15.4 supports unslotted and slotted CSMA/CA, and Time Division Multiple Access (TDMA). Thread only uses unslotted CSMA/CA, which avoids requiring synchronization. While IEEE 802.15.4 provides optional ARQ, data frames are always sent reliably in Thread, aiming to mitigate radio issues (e.g., multipath fading or interference) in a smart home [2].

To support IPv6 over IEEE 802.15.4, Thread employs the IPv6 over Low power Wireless Personal Area Networks (6LoWPAN) adaptation layer [10]. Located between IPv6 and IEEE 802.15.4, 6LoWPAN provides functionality including packet fragmentation (to support the IPv6 MTU requirement, since the basic IEEE 802.15.4 maximum frame size is only 127 bytes), and header compression (to efficiently encode IPv6 and UDP headers).

To enable multihop networking over IEEE 802.15.4, 6LoWPAN defines mesh-under and route-over, by which routing is performed below IPv6 or at the IPv6 layer, respectively. Thread defines a mesh-under, distance vector routing protocol similar to the Routing Information Protocol (RIP) [9]. Thread exploits the small scale and the power supply conditions of a home to offer a simple, full mesh network of up to 32 mains-powered routers. Thread relieves low-power end devices from routing; these devices sleep by default and periodically poll a neighboring router for incoming data. The latter stores data packets intended for the former during sleep intervals. This mechanism, called *polling*, is adopted by Matter as main energy-saving technique.

## IV. Evaluation

This section evaluates the performance of Matter when using Thread and Wi-Fi (i.e., its currently supported IoT technologies) as lower layers. The evaluation focuses on encapsulation overhead (which impacts the next two performance parameters), latency (relevant when a user expects real-time interaction), and energy consumption (crucial for energy-constrained devices).

Experimental results were obtained in an indoor, home environment by using lighting application software over Matter. For Matter over Thread, Nordic nRF52840 DK devices (with 256-kB RAM and 32-bit CPU) were used as endpoint (running *light_bulb* and *light_switch* in home/ncs/v2.1.1/nrf/samples/matter from nRF Connect SDK v2.1.1) and as dongle on a Raspberry Pi 3 B+ router, in a 2-hop network topology. For Matter over Wi-Fi, Espressif ESP32-C3-DevKitC-02 devices (with 400-kB RAM and 32-bit CPU, running *light* and *light_switch* from esp-matter/examples on Espressif's GitHub) were used as endpoints, along a MikroTik hAP ac lite access point (AP); IEEE 802.11n was used. 100 lightbulb turn on and off actions were triggered from a remote switch. Experiments were conducted with default configuration settings, including a transmit power of 0 dBm for Thread and 20 dBm for Wi-Fi, leading to maximum indoor link ranges of ~30 m and ~20 m, respectively. For analysis clarity, distances between neighboring devices were ~2 m with line-of-sight propagation, favoring good network conditions. Network traffic was captured by using a sniffer.

## A. *Encapsulation overhead*

Fig. 3 illustrates the measured encapsulation overhead of each layer for Matter over Thread and over Wi-Fi.

For Matter over Thread, the measured IEEE 802.15.4 overhead is the minimum possible by configuration. However, header compression can be optimized (Fig. 3) by leveraging three techniques: a) full IPv6 address compression (which requires 6LoWPAN context [3], limited to 16 combinations of source and destination IPv6 addresses), b) using 6LoWPAN UDP ports (to reduce a 2-byte port to a 4-bit format), and c) eliding the 2-byte UDP checksum (possible since the Message Layer provides an integrity check via its MIC).

In Matter over Wi-Fi, the overhead from the physical, link and network layers is greater than that of Thread (Fig. 3). Note that IPv6/UDP header compression has not been defined over Wi-Fi, since many Wi-Fi devices are mains-powered.

We expect the measured overheads to match those of a real deployment, except that an optimized Matter over Thread implementation would use the above mentioned header compression techniques b) and c).

## B. *Latency performance*

We next evaluate the latency components associated to turning on/off a lightbulb from a remote switch. In the implementations used, the switch sends a turn-on/off command, the lightbulb sends a response to the switch, and an MRP ACK is finally sent by the switch. The Message Layer payload sizes of these messages are 25 bytes, 33 bytes, and 0 bytes, respectively. Fig. 4 plots endpoint (lightbulb and switch) response times, and Thread router or Wi-Fi AP forwarding times, based on 100 individual measurements (performed on sniffed traffic) for each variable. Frame retries did not occur.

The lightbulb response time comprises processing the received message, turning on/off an LED, creating an action in response and sending it, whereas the switch response time involves a lighter processing (only up to the Message Layer):

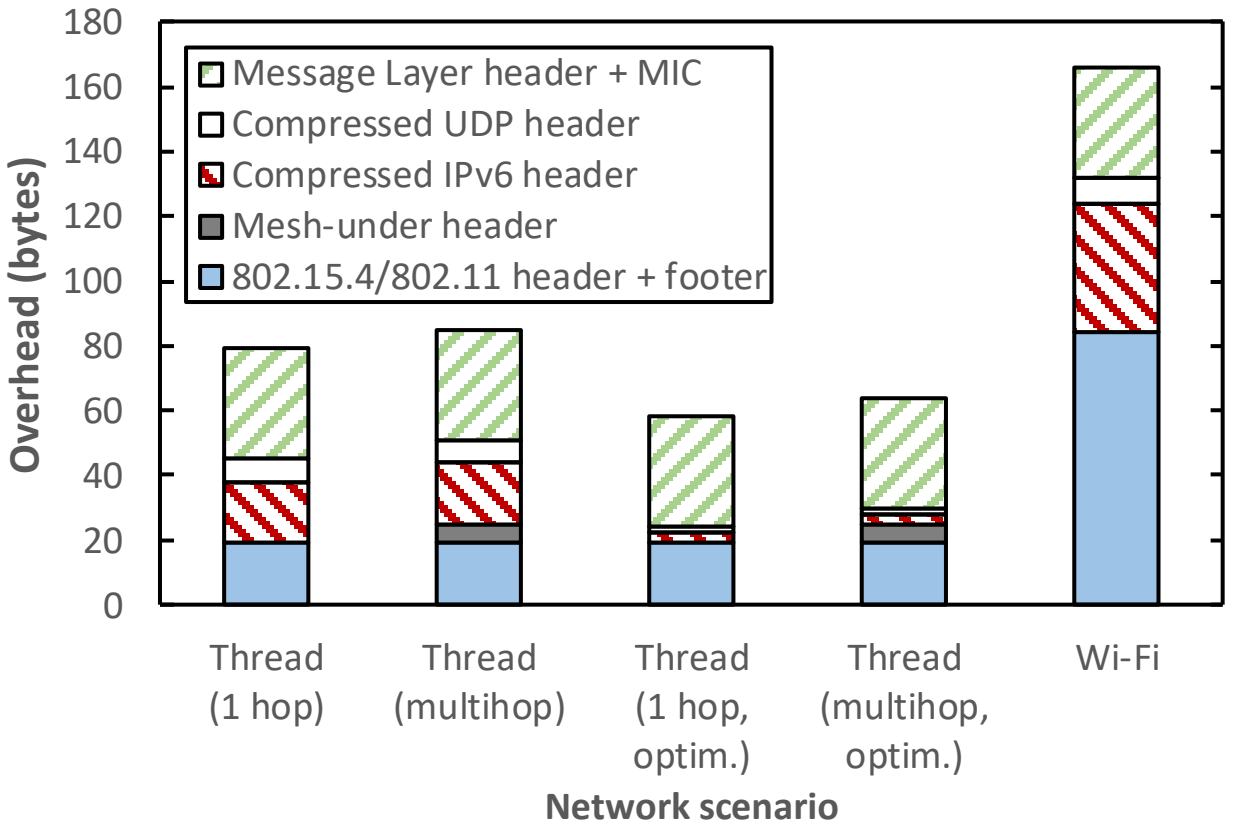

Fig. 3. Encapsulation overhead of Matter over Thread and over Wi-Fi. For comparison, the Message Layer payload size is 25 bytes for the lightbulb turn-on command.

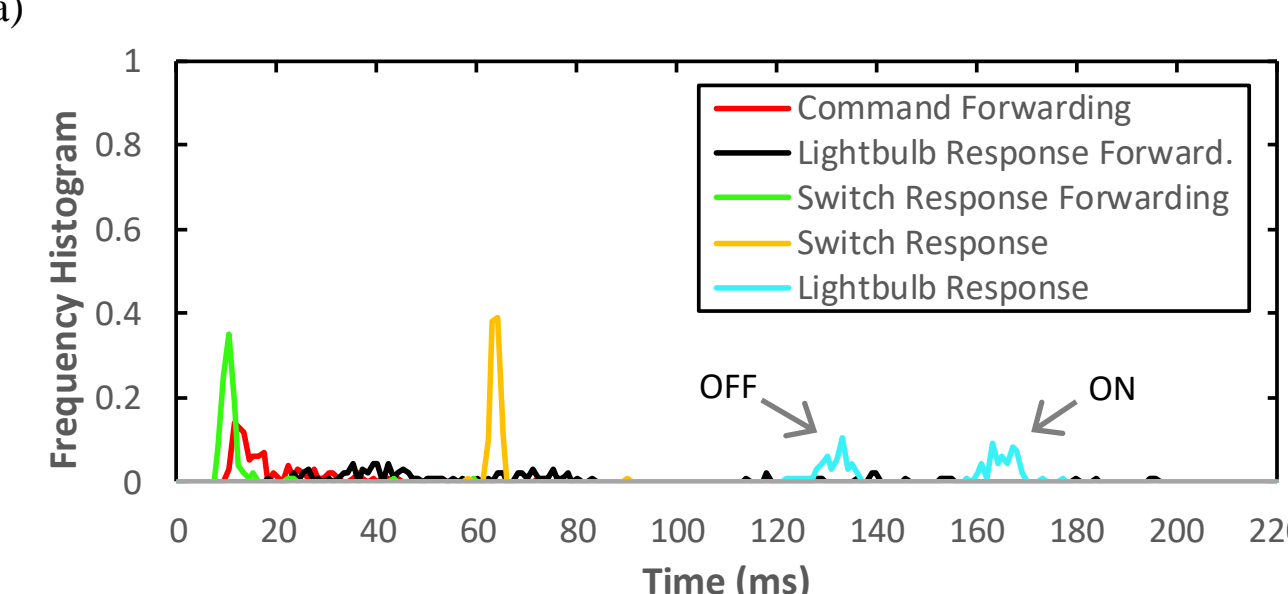

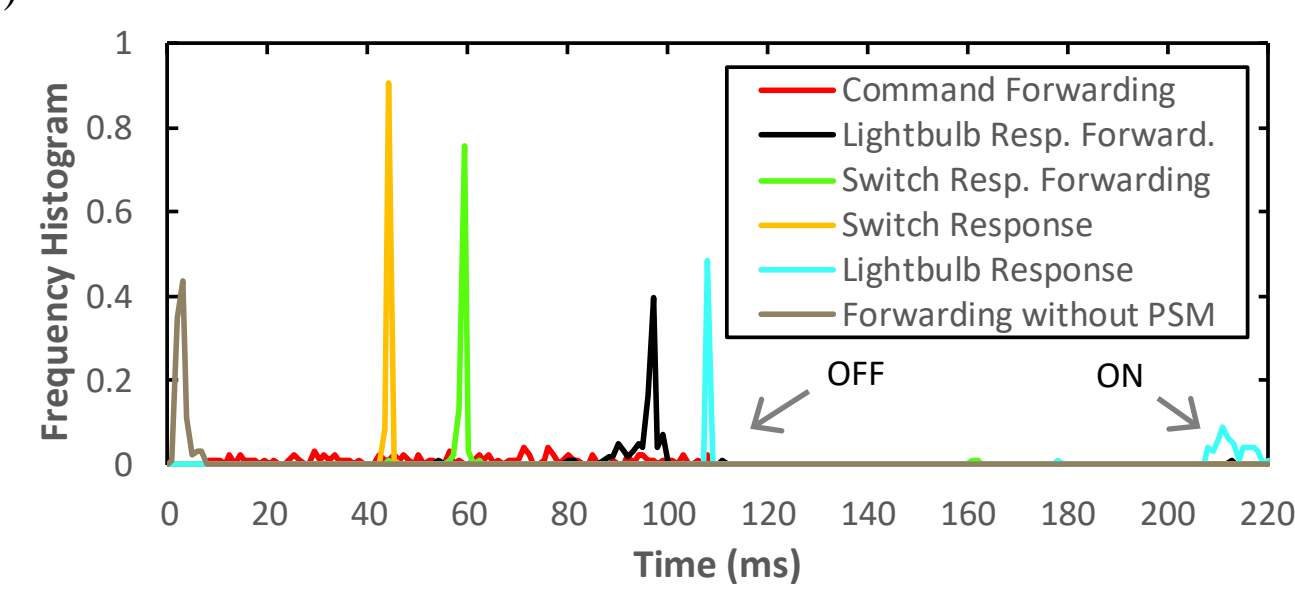

Fig. 4. Frequency histogram of latency components in Matter message exchanges when turning on or off a lightbulb: a) over Thread, b) over Wi-Fi. Handling the physical action of turning on an LED requires more time than turning it off.

generating and sending an MRP ACK. Thus, the former is ~60 to ~170 ms greater than the latter. Over Thread, for messages sent to the lightbulb (i.e., command and switch response), router forwarding time (and its variability) increases with message size, and is generally lower than endpoint response times, since it does not involve functionality atop IPv6. Processing, which is limited by the constrained CPU and RAM resources of the devices, is the main contributor to the mentioned times (note: the theoretical maximum 1-hop transmission time of a lightbulb response is 6.0 ms and 0.3 ms over Thread and Wi-Fi, respectively, including the initial back-off and assuming error-free transmission). However, Thread router forwarding time for the lightbulb response is up to the polling interval (200 ms in the scenario), since the switch is implemented as sleepy device. Over Wi-Fi, both endpoints use IEEE 802.11 Power Saving Mode (PSM), thus they can only receive data after a beacon. Therefore, the AP command forwarding time is uniformly distributed up to the beacon interval (~100 ms). The other AP forwarding times are roughly equal to the time since response transmission by the endpoint until the next beacon transmission. For comparison, Fig. 4.b) includes measured AP forwarding times without PSM. We expect the measured forwarding times to be encountered in a real deployment, whereas response times might be reduced in an optimized Matter implementation.

## C. *Energy performance*

We next evaluate the impact of Matter features and parameters on the lifetime of a battery-operated Thread device that periodically transmits a report (Fig. 5). *Report period* and *poll period* denote the time between the start of two consecutive report and poll message transmissions, respectively. Note that the device sleeps while not reporting or polling. The estimated lifetime and the following list of energy consumption components are based on measurements performed on an nRF52840 DK device with a supply of 3 V, assuming a typical button-cell battery capacity of 230 mAh. Sending a poll message (28 bytes), the shortest Matter message (85 bytes) and the longest non-fragmented Matter message (133 bytes), consumes 52.65 μJ, 76.38 μJ, and 91.26 μJ, respectively, including reception of the subsequent IEEE 802.15.4 ACK in each case. Sleep current is 2.65 μA.

As shown in Fig. 5, for a low poll period (e.g., its default value of 0.3 s), polling dominates energy consumption. Device lifetime increases with the poll period and the report period. However, there exists a trade-off between energy performance and responsiveness of the sleepy device that depends on the poll period. Since packet delivery may be delayed up to a poll period, actuators expected to offer real-time reaction (e.g., lightbulbs, blinds, etc.) need to be mains-powered or require frequent battery replacement. Sleepy devices are appropriate for sensors with mostly unidirectional and infrequent traffic (e.g., sensors measuring temperature, humidity, light, smoke, etc.). Battery lifetime approaches 10 years for poll periods greater than 5 minutes (note: the maximum poll period in Matter is 1 hour). For Matter messages that can be carried in a single IEEE 802.15.4 frame, the impact of message size on battery lifetime becomes negligible for high report periods.

## V. Comparative analysis

This section compares Matter with other prominent smart home protocol stacks, such as ZigBee, Z-Wave, and Bluetooth Mesh (Table 1).

The considered protocol stacks exhibit similarities regarding the functionality supported atop the network layer, such as E2E encryption and authentication, and -except for Z-Wave- optional E2E reliability with ARQ, and client-server operation.

Significant differences across solutions arise at the network layer, with diverse approaches for E2E packet delivery. Matter is the only solution designed for interoperability (i.e., originally based on IP). In contrast with the other protocol stacks, Matter allows to integrate devices of various underlying technologies and provides IP-based Internet access.

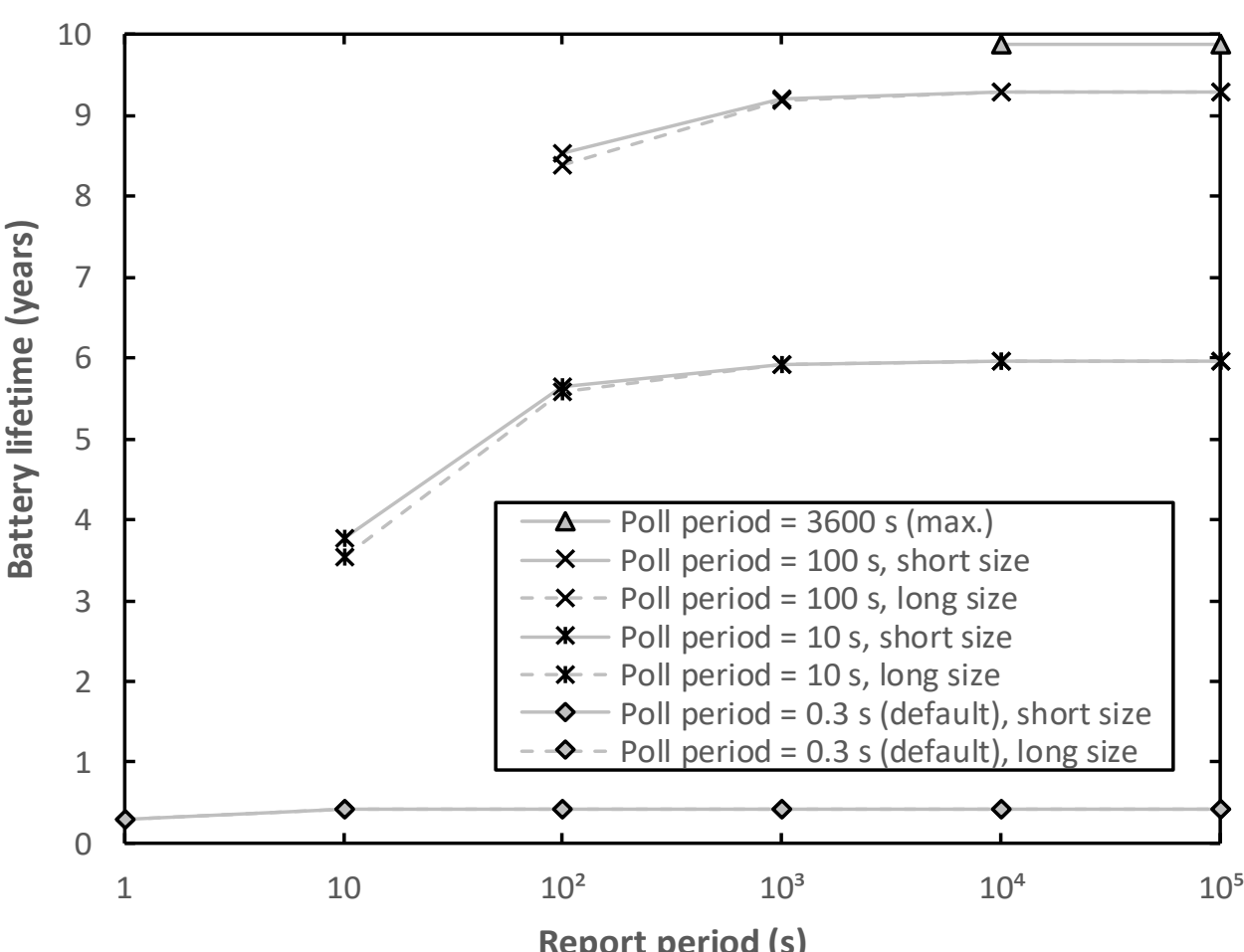

Fig. 5. Impact of Matter on the maximum battery lifetime of the considered Thread device, for different parameter settings, and for the shortest and the longest non-fragmented Matter message size.

However, Matter also exhibits the greatest protocol encapsulation overhead (with a significant contribution from the Message Layer, Fig. 3). Thus, latency, energy, processing and memory become penalized, especially, in resource-constrained networks. For the latter, ZigBee, Z-Wave or Bluetooth Mesh offer more lightweight approaches.

Finally, since Wi-Fi was not originally intended for constrained environments, it offers higher bitrate and frame size, and Matter over Wi-Fi does not require an IP adaptation layer or multihop functionality.

## VI. The Cost of Interoperability

We next analyze development and consumer costs for the two main approaches in smart home interoperability: a universal solution like Matter, and hub-based solutions. We assume $N$ smart home devices, each one using a different communication solution.

With Matter, hub development cost is 0, since a hub is not required; device communication software needs to be developed only for one solution (i.e., Matter). For the customer, hub acquisition cost is 0, and device acquisition cost is constant with $N$. Otherwise, at least one hub supporting all communication solutions is needed (e.g., the approach in Home Assistant) and, at most, one hub is required for each

TABLE 1

Main Features of Smart Home Protocol Stacks. 'ZSE' Stands for ZigBee Smart Energy 2.0 or Subsequent.

<table>
<tr><th></th><th>Matter over Wi-Fi</th><th>Matter over Thread</th><th>ZigBee</th><th>Z-Wave</th><th>Bluetooth Mesh</th></tr>
<tr><td>Bitrate (Mbps)</td><td>2 – 9600 (max.)</td><td>0.25</td><td>0.02, 0.04, 0.25</td><td>0.0096, 0.04, 0.1</td><td>0.125, 0.5, 1, 2</td></tr>
<tr><td>L2 reliability</td><td>ARQ (always)</td><td>ARQ (always)</td><td>ARQ (typically)</td><td>ARQ (optional)</td><td>Each transm. is done thrice</td></tr>
<tr><td>Max. L2 pay. size (bytes)</td><td>2304</td><td>102</td><td>102</td><td>130</td><td>255 (from 5.0)</td></tr>
<tr><td>Low-power support</td><td colspan="2">Polling</td><td>Polling</td><td>Channel sampling[1]</td><td>Polling</td></tr>
<tr><td>IPv6 support</td><td colspan="2">Yes</td><td>Only in ZSE</td><td>No</td><td>No</td></tr>
<tr><td>Adaptation layer</td><td>No</td><td>Mesh-Under 6LoWPAN</td><td>Route-Over 6LoWPAN (ZSE)</td><td>N.A.</td><td>N.A.</td></tr>
<tr><td>E2E packet delivery</td><td>Single-hop (IP perspective)</td><td>Routing (proactive, distance vector)</td><td>Routing (reactive, dist. vect.); RPL (ZSE)</td><td>Routing (proactive, link state)</td><td>Controlled flooding</td></tr>
<tr><td>E2E reliability</td><td colspan="2">TCP, MRP (optional)</td><td>ARQ (optional); TCP/CoAP (ZSE)</td><td>No</td><td>ARQ (optional)</td></tr>
<tr><td>App. model</td><td colspan="2">Client-server</td><td>Client-server</td><td>Master-slave</td><td>Client-server</td></tr>
<tr><td>E2E encrypt. and auth.</td><td colspan="2">Yes (16-byte MIC)</td><td>Yes (4-byte MIC)</td><td>Yes (8-byte MIC)</td><td>Yes (4-, 8-byte MIC)</td></tr>
<tr><td>Encaps. overhead (bytes)</td><td>166</td><td>79</td><td>36</td><td>21</td><td>29</td></tr>
<tr><td>Organization</td><td colspan="2">CSA</td><td>CSA</td><td>Z-Wave Alliance</td><td>Bluetooth SIG</td></tr>
</table>

[1]The device periodically samples the channel for incoming data.

pair of devices. Hub development and acquisition costs increase with *N* between linearly and quadratically. Device development cost grows linearly with *N*, and device acquisition cost increases with *N*.

## VII. Future Directions

This section discusses future directions for Matter.

### A. *Simplifying the application layer*

Future Matter versions may benefit from simplifying application-layer formats and operation, to reduce complexity and delay. Unused Message Layer header field values, and long session/message identifier and MIC sizes, offer optimization opportunities. However, the trade-off with protocol extensibility, error detection, and security needs careful consideration.

### B. *Reducing Message Layer header overhead*

Static Context Header Compression (SCHC) [11] may be used to reduce the current Message Layer header overhead. SCHC exploits context shared by the compressor and the decompressor, based on a priori knowledge of header field values. SCHC may compress an 18-byte Message Layer header down to 2 bytes.

### C. *Cross-technology communication*

A promising avenue for energy and bandwidth improvement in Matter is Cross-Technology Communication (CTC). In CTC, devices equipped with a wake-up radio receiver may communicate directly (i.e., without a router), despite using different technologies (e.g., Wi-Fi and Thread). SCHC may be used as adaptation layer to support IPv6 over CTC [12].

### D. *Additional technologies*

Being based on IPv6, Matter is expected to incorporate additional lower layer technologies in the future.

BLE is a clear candidate, since it is already supported for commissioning in Matter. The omnipresence of BLE in smartphones would allow them to become user devices for Matter networks. Furthermore, BLE multihop extensions can address smart home coverage issues.

LoRaWAN is another promising candidate low-power technology for Matter. LoRaWAN popularity is increasing and now it supports IPv6. However, some LoRaWAN bitrates (e.g., below 1 kbit/s) are unsuitable for real-time applications with humans in the loop.

### E. *Expanding beyond the smart home domain*

If Matter succeeds in smart homes, a natural next step for it would be expanding to other application domains. For the latter, new clusters would need to be defined. However, since Matter's Data Model is based on the ZCL, and various ZigBee application profiles exist, such expansion would require a limited effort. Furthermore, existing clusters may be partially reused for other domains, such as building automation, smart health, smart lighting, and smart energy.

## VIII. Conclusions

Thanks to its IPv6-centric approach and its new application-layer protocol, Matter may become a universal communication solution for smart homes. Furthermore, Matter has the potential to achieve greater performance, extend its set of lower layer technologies, and expand beyond the smart home.

## Acknowledgment

This work was partly supported by the Spanish Government (PID2019-106808RAI00, PID2023-146378NB-I00), and by Generalitat de Catalunya (2021 SGR 00330). The experiments dataset is accessible (https://doi.org/10.34810/data1813).

## Biographies

Saeid Madadi-Barough is a PhD student at Universitat Politècnica de Catalunya (UPC). His research focuses on IoT.

Pau Ruiz-Blanco obtained his BSc in Network Engineering from UPC. He focuses on IoT and software development.

Jiadeng Lin obtained his BSc in Telecommunication Systems Engineering from UPC. He focuses on IoT and network operation.

Rafael Vidal is an Associate Professor at UPC. His research interests include wireless networks and IoT.

Carles Gomez is a Full Professor at UPC. His research focuses on IoT.